# How special are the dynamics of deep eutectic solvents? A Look at the Prototypical Case of Ethaline

Mohammad Nadim Kamar[a], Armin Mozhdehei[a], Basma Dupont[a], Ronan Lefort[a], Alain Moréac[a], Jacques Ollivier[b], Markus Appel[b], and Denis Morineau[a*]

[a]Institute of Physics of Rennes, CNRS-University of Rennes, UMR 6251, F-35042 Rennes, France

[b]Institut Laue-Langevin, 71 avenue des Martyrs, F-38000 Grenoble, France

* denis.morineau@univ-rennes.fr

ABSTRACT We investigated the molecular dynamics of the prototypical deep eutectic solvent (DES) ethaline. We disentangled the different motions of its two constituents, namely choline chloride and ethylene glycol on a spatio-temporal range that extends from sub-nanometer to micrometer distances and from picosecond to millisecond times. This was achieved by a combination of pulsed-field-gradient NMR, time-of-flight, and backscattering quasielastic neutron scattering experiments with isotopically labelled samples. On the micrometer scale, we observe that the translational motions of the two DES constituents obey classical hydrodynamics, with distinct diffusivities that reflect their different hydrodynamic radii. This is no longer valid at the nanometer-scale, where the two DES components present similar short-ranged diffusivities, which indicates a significant effect of their supramolecular association. The sub-nanometer scale motions include jumps that precede Fickian diffusion, and localized dynamics that precede the breaking of the transient cage formed by neighboring molecules. Therein, the spatial amplitude of the localized motions mirrors their different molecular sizes and chemical structures, while their respective correlation times contrast with observations made for other choline-based DES such as glyceline. This result underlines the importance of more subtle effects, such as the different H-bond propensities of the polyol donor, and demonstrate the difficulty to anticipate the nanoscale dynamic behavior of DES from the knowledge of their macroscopic properties.

## INTRODUCTION

Over the past decade, deep eutectic solvents (DESs) have emerged as a promising class of solvents that demonstrate wide-ranging applicability in various industrial processes, including electrode metal deposition,[1–5] catalysis,[5] nanoparticle and nanotube synthesis,[5] drug transport, and carbon dioxide capture.[6] They have opened new perspectives towards the design of eco-friendly and cost-effective alternatives to ionic liquids (ILs).[7]

Fundamentally, DESs also present exciting physicochemical properties that are not fully understood yet. Their phase diagrams show reduction of the eutectic melting point that exceeds the thermodynamic predictions based on the ideal mixing assumption.[8,9] This phenomenon indicates that specific interactions, such as electrostatic forces and H-bonds between the DES components plays a crucial role. Consequently, the structure of DESs is more complex than classical molecular solvents. This was first confirmed for reline, a DES formed by choline chloride and urea, by wide-angle neutron scattering experiments that revealed the formation of stoichiometric supramolecular ionic complexes.[10] The structural microheterogeneous nature of DESs has been further observed by different methods, including small-angle neutron scattering,[11] X-ray diffraction, and molecular dynamics simulations.[12–14]

The dynamics of DESs present intriguing features, which have been observed using different techniques such as nuclear magnetic resonance (NMR),[15–18] dielectric spectroscopy (DS),[19–21] molecular dynamics (MD) simulations,[22] and quasielastic neutron scattering (QENS).[23,24] Diverse observations were reported depending on the studied DES system and the applied method, which appear somehow contradicting and demonstrate the complexity of DES dynamics. This is the case for prototypical solvents that belong to the same class of choline chloride (ChCl)-based DES, including reline, ethaline and glyceline, which are formed by mixing ChCl with urea, ethylene glycol, and glycerol as hydrogen bond donor (HBD), respectively. The as-formed DES mixtures have smaller viscosity and faster dynamics than the corresponding pure HBD for glyceline, but not for ethaline. [20,21] For glyceline and ethaline, comparable temperature dependences of the viscosity, dipolar relaxation and ionic conductivity were reported, indicating that Stokes-Einstein, Debye-Stokes-Einstein and Walden rules are obeyed, which is unlike reline that presents enhanced ionic mobility and fractional Walden rule.[19,20] The decoupling between the dynamics of the different constituents of the DES was actually reported from DS experiments by Spittle et al.[21] who attributed a secondary slow relaxation process of glyceline to motions of ChCl that were coupled to ion dynamics but decoupled from the faster main structural relaxation driven by glycerol dynamics. In this context, chemically sensitive methods, such as QENS and NMR, provide unique opportunity to disentangle the different components. The QENS method was applied for glyceline by Wagel et al.,[23] reaching the opposite conclusion that localized diffusive motions of ChCl are effectively larger than those of glycerol. This result is also supported by the deuteron NMR study. [17,18] Strikingly, this systematic deuteron NMR work on the three ChCl-based DESs confirms that the rotational dynamics of ChCl is faster than the HBD counterpart for glyceline and for reline, but it unravels the opposite behavior for ethaline.

In the rapidly evolving field of DES, an important question concerns the potential universality of their behavior, and whether features observed for given prototypes can be generalized to an entire class of DES. Accordingly, the unusual dynamics of ethaline, compared to glyceline and reline is outstanding. As such, it requires further in-depth investigation, which is the purpose of the present work. Along with NMR, QENS with isotopic labelling offers unique capability to perform H-tagged molecules dynamics studies that benefit from the high incoherent scattering cross-section of hydrogen.[25,26] We performed QENS experiments for two different selectively deuterated ethaline samples in order to highlight the relative contributions from the choline chloride and ethylene glycol components of the DES mixture. Combining NMR and QENS results, we achieved an unprecedented view on the molecular dynamics of ethaline, which revealed couplings between the two DES constituents and deviations from classical hydrodynamics in magnitude that varies with the observation length scale. These findings provide a better understanding of the complex microscopic dynamics of DES, which can impact their adaptation to specific applications depending on their relevant time and length scales.

## METHODS

**Samples**

Fully hydrogenated anhydrous ethylene glycol (EG) (99.8% purity) and choline chloride (ChCl) (> 99% purity) were procured from Sigma-Aldrich. Partially deuterated ethylene glycol (1.1.2.2 $D_4$, 99% D) and choline chloride (trimethyl $D_9$, 98% D) were purchased from Eurisotop. These chemicals were directly employed to prepare DES without additional purification. Ethaline mixtures were prepared at different compositions that correspond to one mole of choline chloride combined with $n$ moles of ethylene glycol, with $n$ = 2 or 4. For NMR, fully hydrogenated ethaline samples were prepared for two specific molar ratios 1:2 and 1:4. For QENS, two ethaline samples were prepared for specific molar ratio 1:4 but using two different isotopic compositions, so that one of the two constituent is deuterated and the other fully hydrogenated. They are later denoted ChCl(H)/EG($D_4$) and ChCl($D_9$)/EG(H). The mixtures were prepared in a glove box and mechanically agitated at 60°C for 30 minutes until a clear and homogeneous liquid phase formed.

Deuteration is a crucial technique in neutron scattering studies, particularly in quasi-elastic neutron scattering (QENS), to selectively label a specific component within a mixture by tuning its incoherent scattering contribution. Hydrogen ($^1$H) possesses a high incoherent scattering cross-section ($\sigma_{inc}$= 80.27 barns), which can dominate the neutron signal and limit structural or dynamical information from other components. By replacing hydrogen with deuterium ($^2$H), which has a significantly lower incoherent cross-section ($\sigma_{inc}$= 2.05 barns), the total incoherent contribution of the targeted molecule is significantly reduced, allowing enhanced contrast in the neutron signal from H-tagged species. The coherent and incoherent scattering cross-section of each type of isotopically labelled molecule is provided in Table S1. The relative contribution of each constituent of ethaline for the studied specific molar ratios 1:4 is illustrated in Fig. 1.

Accordingly, for ChCl(H)/EG($D_4$) the neutron cross section is dominated by the incoherent scattering from the cholinium cation, with a secondary contribution from EG (about 1/3). The relative contributions are inverted for ChCl ($D_9$)/EG (H), where ¾ of the total cross section arise from the incoherent scattering of EG. It should also be noted that the contribution from coherent scattering to the total scattering intensity is in general quite small (7%-12%). However, due to its $Q$-dependence, this needed to be carefully addressed. Indeed, the contribution of coherent scattering is expected to be modulated in $Q$ in a similar way to the static structure factor, which presents vanishing amplitude in the limit of small $Q$, and maximum intensity in the region of the main diffraction peak at about 1.5Å-1. We observed that the static and elastic structure factors presented featureless $Q$-dependence, as expected for incoherent scattering intensity, which actually ruled out any significant (i.e., measurable) contamination from coherent scattering, as shown in Fig. S1 and S2.

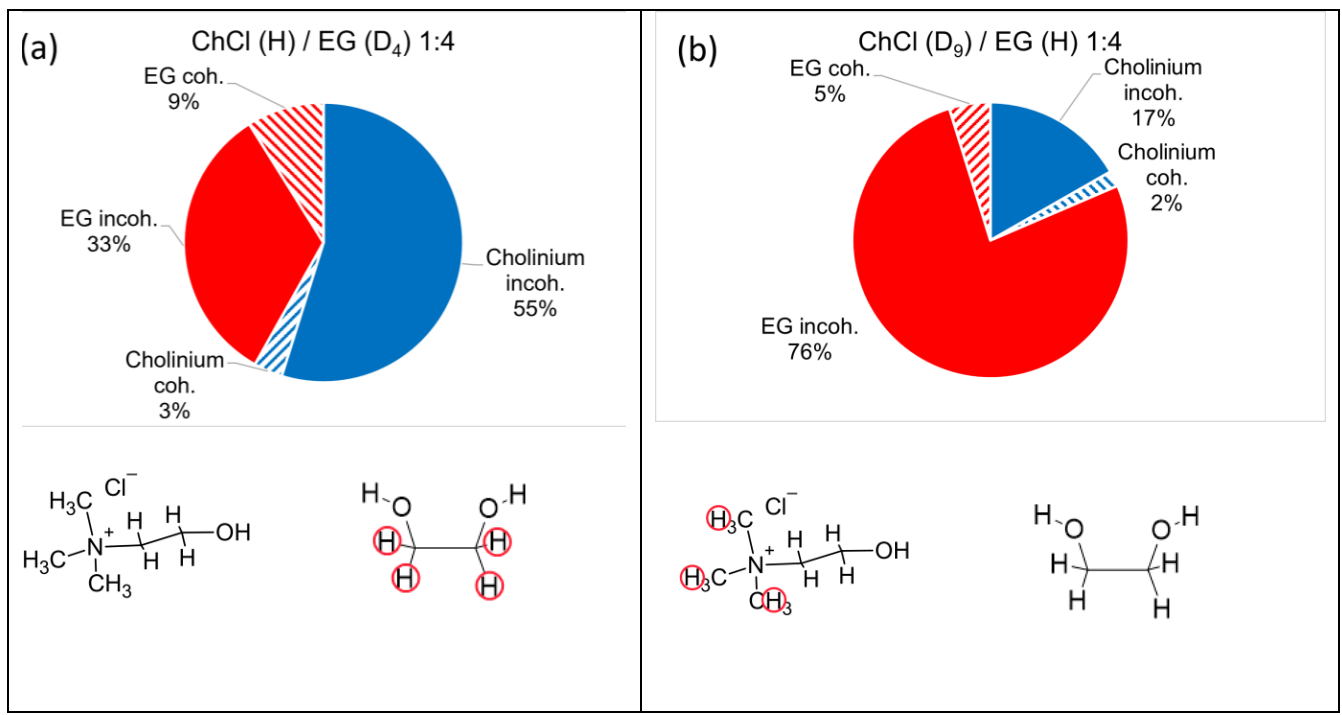

**FIG. 1.** Relative contributions from the different molecular constituents to the total neutron scattering cross section of ethaline (1:4) comprising (a) partially deuterated EG and (b) partially deuterated ChCl. The molecular structure of the corresponding mixtures is illustrated in lower panel with deuterated sites indicated by red circles.

**QENS experiments**

Quasielastic neutron scattering (QENS)[25] experiments were performed using two spectrometers with distinct energy resolutions at the Institut Laue-Langevin (ILL) in Grenoble, France.[27] The disk chopper time-of-flight spectrometer, IN5B, was employed with an incident wavelength of 4.9 Å. Under this configuration, the energy resolution ($\delta E$) at the elastic peak was approximately 80 µeV (FWHM), corresponding to a timescale ($t = \hbar/\delta E$) of roughly 10 ps. The retained quasielastic signal for data evaluation spanned an energy range ($E = \hbar\omega$) between −1 meV and 1 meV and a $Q$-range from 0.2 to 2.1 Å$^{-1}$. The IN16B backscattering spectrometer provided a higher resolution (by about a factor of 100). Unpolished Si(111) monochromator and analyzers were used to set the incident wavelength of 6.27 Å, an energy resolution of $\delta E$ = 0.75 μeV and a resulting timescale of around 1 ns. The instrument was utilized for elastic and inelastic fixed window scans (FWS) at energy transfers of elastic (0 μeV), 3 μeV and 6 μeV. [28,29]

By combining neutron time-of-flight (ToF) spectroscopy with high-resolution backscattering (BS) spectroscopy, the dynamics of ethaline DES were explored over a broad timescale. A cryofurnace was used in order to regulate the sample temperature. For IN5B, measurements were performed after achieving thermal equilibrium at three selected temperatures 300 K, 325 K, and 350 K, while for IN16B, a continuous temperature ramp was applied on cooling from 350 K to 10 K. Both isotopic DES mixtures and pure ethylene glycol were studied.

The raw data were normalized to the incoming neutron flux and corrected for detector efficiency using vanadium as a reference. Background contributions from the spectrometer and empty containers were subtracted, and the experimental intensity was transformed into the dynamic structure factor, $S(Q,\omega)$. Data corrections and analysis were performed using the MANTID software package,[30] Data fitting of IFWS was carried out with MANTID scripts, while fitting of $S(Q,\omega)$ in the frequency domain was conducted with the QENSH program provided by the Laboratoire Léon Brillouin (LLB, Saclay, France).

## RESULTS

**PFG NMR measurements**

The self-diffusion coefficients measured by PFG NMR are measured over molecular displacements that are on the micrometer scale. Hence, they provide direct information about the liquid translational dynamics at the macroscopic (i.e., hydrodynamic) limit, hence a wave vector Q range much lower than in the QENS experiments.

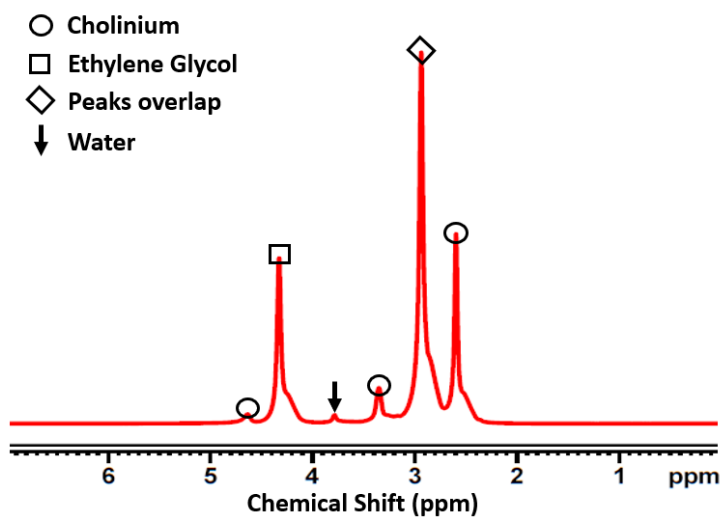

**FIG. 2**. $^1$H NMR spectrum of ethaline (1:4) at 298 K (TMS as reference).

The $^1$H NMR spectrum of ethaline at 298 K, highlighting the resonances of ethylene glycol ($\delta$ = 3.05 ppm for the –CH$_2$– resonance and $\delta$ = 4.35 ppm for the –OH resonance) along with those of the cholinium cation. Notably, the –CH$_2$– resonance of ethylene glycol at $\delta$ = 3.05 ppm overlaps with one of the –CH$_2$– resonances of the cholinium cation. Also, one observes a slight broadening of the peak linewidths, which is primarily attributed to the intrinsic properties of DESs. Due to their high viscosities, DESs exhibit strong inter- and intra-molecular dipolar interactions, which contribute to the broadening effect observed in the NMR line shapes. Importantly, the presence of distinct resonances attributed respectively to EG and ChCl, allows measuring the diffusivity of each component of the mixture using the fully hydrogenated systems.

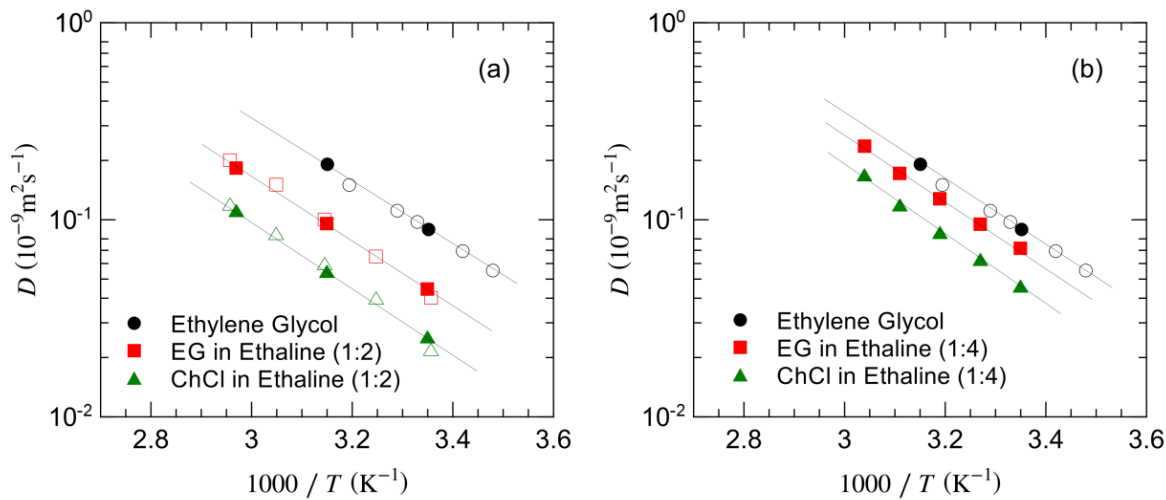

**FIG. 3.** Arrhenius representation of the self-diffusion coefficient measured by PFG NMR for pure EG (black circle), EG in ethaline (red square), and ChCl in ethaline (green triangle). Two different molar compositions of ethaline are compared (1:2) (left panel) and (1:4) (right panel). Comparison with existing data for pure EG (small open circle) (after Spees et al. [31]), and for the two constituents of ethaline (1:2) (after Alfurayj et al. [32]).

For all studied samples, we actually observed that the NMR signal attenuation conforms with the model of simple Fick diffusion, as illustrated in Fig. S3. The obtained self-diffusion coefficients are illustrated in Fig. 3(a) for ethaline (1:2) and in Fig. 3(b) for ethaline (1:4), as well as for pure EG. Our measurements for ethaline (1:2) and pure EG are in excellent agreement with previously reported data [23,24,31,31]. Since no equivalent results exist in the literature for ethaline (1:4) our additional measurements provide useful information about the impact of the molar composition on the dynamics of ethaline. Also, they are required for a comparison with those evaluated by QENS for the same composition.

The self-diffusion coefficient of EG systematically decreases when going from pure EG, to ethaline (1:4) and finally to ethaline (1:2). This trend is consistent with the observed higher viscosity of ethaline DES compared to pure EG, and also with the monotonous viscosity variations as a function of composition in a range of molar fraction encompassing the two studied mixtures, as reported by Shi et al. [33,34]. The second salient feature is the systematic slower diffusion of ChCl compared to EG for the two mixtures. This phenomenon can be attributed to the larger size of ChCl compared to EG. On the temperature range studied, the ratio between the self-diffusion coefficient of EG and ChCl is $\frac{D_{EG}}{D_{ChCl}} \approx 1.7$ for ethaline (1:2) and $\frac{D_{EG}}{D_{ChCl}} \approx 1.5$ for ethaline (1:4), which is compatible with the ratio between the two molecular sizes $R_{ChCl} = 3.29$ Å and $R_{EG} = 1.5$-$2$ Å [15,35]. According to the Einstein-Stokes equation ($D = \frac{k_B T}{6\pi \eta r}$), one expects that the diffusion coefficients vary in proportion of both the inverse viscosity ($\eta$) and the inverse molecule size ($r$). As a whole, our PFG NMR measurements indicate that in this range of temperature and composition, ethaline translational dynamics agree well with classical hydrodynamics. We can conclude that the diffusive behavior of ethaline at the macroscopic limit is rather typical of a normal simple liquid and does not exhibit anomalous trends. Given the microheterogeneous nature of many DESs, this conclusion must of course be carefully reconsidered at a smaller (i.e., nanometric) scale, as discussed in the QENS part of this work.

**QENS measurements on the IN5B neutron time-of-flight spectrometer**

On IN5B, the QENS spectra measured for pure EG and for ethaline (1:4) DES with two different isotopic compositions are illustrated in Fig. 4 at the three studied temperatures and for the selected value of the transfer of momentum $Q = 1.5$ Å$^{-1}$. We observed a quasielastic broadening that covered a typical energy range from about 0.1 to 1 meV on IN5B. Given the dominant contribution from the incoherent scattering cross section of hydrogen atoms, this quasielastic signal is attributed to the self-part of the dynamic structure factor $S(Q,\omega)$. Thus, it provides information on the hydrogen particles dynamics of the system molecules. A continuous sharpening of the quasielastic peak was observed on decreasing the temperature from 350 K to 298 K. On a qualitative level, the width of the quasielastic scattering is inversely proportional to the typical timescale of motion of particles. Therefore, this sharpening indicates that the dynamics slows down gradually on cooling. The absence of a sharp increase of the elastic signal confirms that ethaline remained liquid on the entire studied temperature range. Compared to pure EG (Fig. 4(a)), a slight sharpening of the spectra can be witnessed for the DES (Fig. 4(b) and Fig. 4(c)), which indicated slower dynamics.

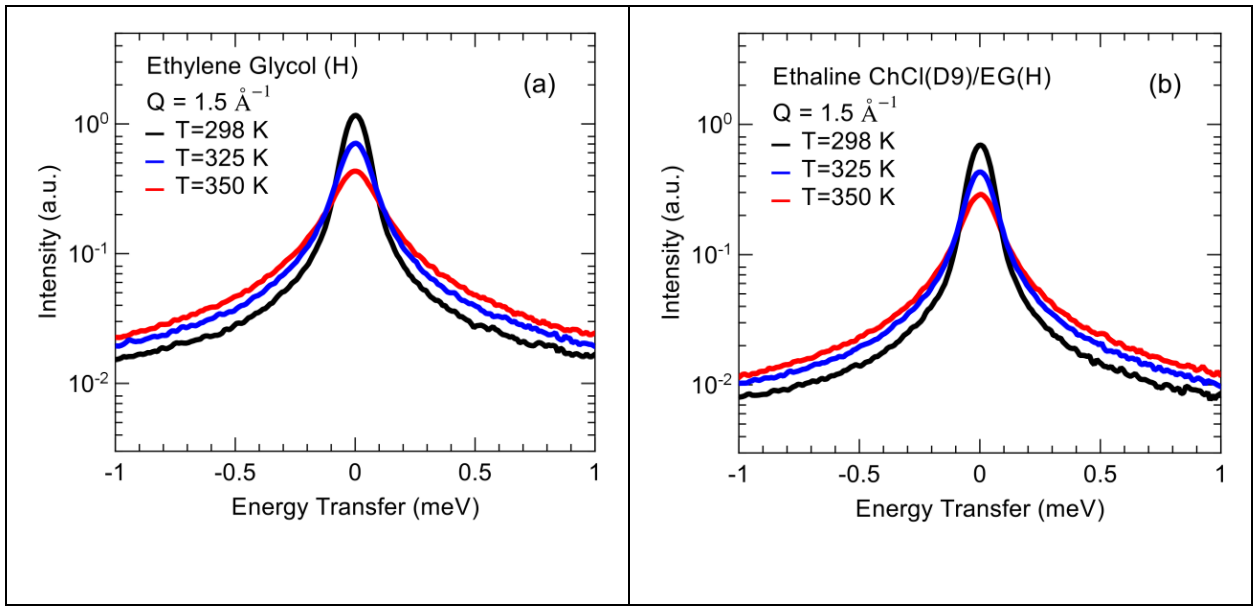

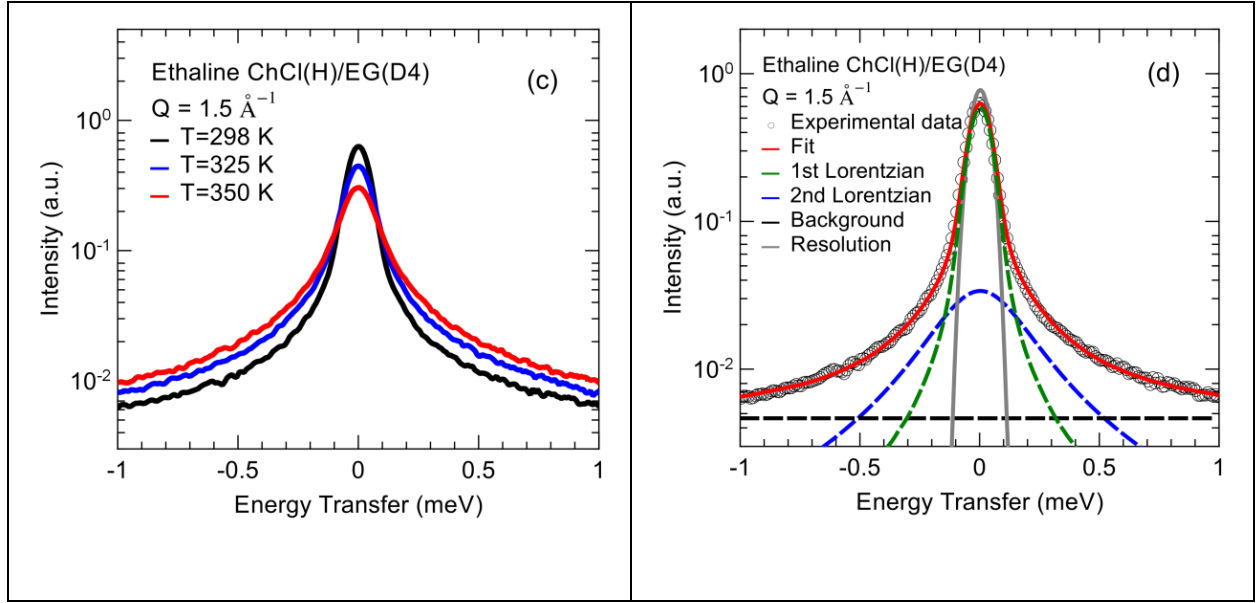

**FIG. 4.** Temperature dependence of the scattering intensity of (a) ethylene glycol, (b) ChCl(D9)/EG(H) and (c) ChCl(H)/EG(D4), QENS spectra measured on IN5B at $Q = 1.5$ Å$^{-1}$ (d). Fit to experimental spectrum (circle) of a model (solid red line) comprising a sharp Lorentzian (green dashed line), second broad Lorentzian (blue dashed line) and flat background (black dashed line).

To extract quantitative description, all spectra were fitted individually at each $Q$ with a model comprising a quasielastic contribution, which was approximated by a sum of two Lorentzian functions convoluted by the instrumental resolution function $R(Q, \omega)$, which was measured with a vanadium sample. The fact that a minimum of two Lorentzian functions were needed to reproduce the experimental data (see Fig. 4(d)) suggests that two dynamical processes were active in the frequency window covered by IN5B. As classically done for liquids, we assigned the two different motions to localized motion $S_L(Q,\omega) = A_L(Q)\delta(\omega) + (1 - A_L(Q)) L_L(Q,\omega)$ and translational diffusion $S_T(Q,\omega) = L_T(Q,\omega)$ where $L_X(Q,\omega)$ stands for a Lorentzian function of width (HWHM) $\Gamma_X$. Unlike translational diffusion, the dynamic structure factor of localized motion contains an elastic term $A_L(Q)$, that mostly arises from the elastic incoherent structure factor (EISF) and reflects the spatial trajectory of the local dynamic. Assuming that the two types of motions are actually independent, the dynamic structure factor writes

$$S(Q, \omega) = K \left[ S_T(Q, \omega) \otimes S_L(Q, \omega) \right] \qquad (1)$$

where $\otimes$ indicates a convolution in $\omega$. Here, $K$ is a scaling factor that accounts for the reduction in scattering intensity caused by inelastic vibrational modes, commonly described by the Debye-Waller factor. The model ultimately writes as

$$S(Q, \omega) = K[A_L(Q) L_T(Q, \omega) + (1 - A_L(Q)) L_{L+T}(Q, \omega)] + B(Q) \qquad (2)$$

where $B(Q)$ is a flat background that needed to be added to account for broader contributions from quasi and inelastic dynamics that significantly exceed the fitted energy range. The intensity and the width of the two Lorentzian functions were determined independently at each $Q$. The fit revealed that the linewidth of the second (broadest) Lorentzian was barely varying with $Q$ (see Fig. S4). Contrariwise, the sharpest Lorentzian was dispersive, which is in line with their assignment to localized and translational motions, respectively.

**Fixed Window Scans measurements on the IN16B neutron backscattering spectrometer**

The elastic and inelastic fixed-window scans (FWS) acquired on IN16B in cooling from 350 K to 10 K and summed over all transfers of momentum $Q$ are illustrated in Fig. 5 for the same three samples as shown for IN5B.

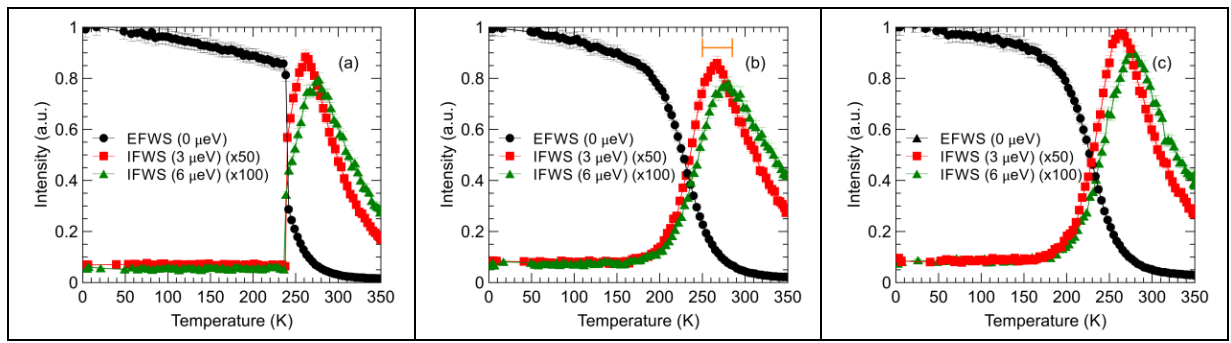

**FIG 5.** Elastic and Inelastic Fixed Window Scans normalized to maximum elastic intensity and summed over all values of transfer of momentum, measured on IN16B for (a) ethylene glycol, (b) ethaline ChCl(D9)/EG(H) and, (c) ethaline ChCl(H)/EG(D4). Intensity obtained at elastic position (black circle) and for two different energy off-sets 3 µeV (red square), and 6 µeV (green triangle). Scaling factors (50x and 100x) have been applied to IWFS for better clarity. The region around the maximum peak in (c) ranges from 250 K to 285K (orange segment).

The slowdown of dynamics on cooling was demonstrated by the increase of elastic intensity. For EG, the liquid supercooling was interrupted by a sudden crystallization that is illustrated in Fig. 5(a) by a discontinuous jump of elastic intensity ($Tc \approx 240$ K). Unlike EG, the increase of elastic intensity was gradual for both DESs. This confirms the absence of crystallization and the glassforming ability of ethaline ($Tg \approx 157$ K) as already reported by calorimetry and dielectric spectroscopy. [8,36]

In the meantime, the inelastic fixed window scans (IFWS) measured at 3 µeV (resp. 6 µeV) presented a single peak that covers the temperature range from about 240 K (resp. 250 K) to 330 K (350 K) at half maximum intensity. The peak maximum positions, which are centered around 270 – 280 K in Fig. 5 actually shift to a higher temperature on increasing the energy offset, and as the value of the transfer of momentum decreased as illustrated in Fig. 6 for ChCl(H)/EG(D4). This evolution reveals the dispersive nature of the associated dynamics, which is attributed to the $Q$-dependent onset of translational diffusion dynamics within the dynamical range fixed by the selected energy off-sets. In fact, the second broader quasielastic component seen on IN5B spectra that is related to local molecular motions considerably exceeds the energy range covered by IN16B. Therefore, a simple model expressed by Eq. (3) was fitted to the IFWSs. It comprises a Lorentzian term $L_T$ related to translational diffusion and a background $BG(Q)$ accounting for faster local motions. The IWFS measured at two offset energies $\hbar\omega_{offset}$ were fitted simultaneously using the same set of parameters, but each $Q$-value was considered independently.

$$I^{fit}_{IFWS}(Q,T) = A(Q)L_T(Q,\omega_{offset}) + BG(Q) \qquad (3)$$

The temperature dependence of the linewidth $\Gamma_T(Q,T)$ of the Lorentzian function $L_T$ was assigned to obey the Arrhenius law, with a same activation energy parameter $E_a$ for all $Q$'s and offset energies, according to Eq. (4)

$$\Gamma_T(Q,T) = \Gamma_0(Q)exp\left(\frac{-E_a}{k_B T}\right) \qquad (4)$$

Unlike IN5B experiments, the linewidths $\Gamma_T(Q,T)$ are not obtained at different specific temperatures on IN16B. In fact, the prefactor $\Gamma_0(Q)$ and the activation energy $E_a$ obtained from the modelling of IN16B intensities reflect the Arrhenian dependence of $\Gamma_T(Q,T)$ that best reproduce the entire evolution of the IFWS on the studied temperature range. That being said, it must also be recognized that the reliability of the quantities obtained is optimal in the region where the dynamics of the system best corresponds to the selected energy transfer, which is located around the peak maximum. In Fig. 5(b), we have illustrated this region by a segment extending from 250

K to 285 K. This corresponds to a rather strict definition of the temperature range where the DES dynamics are accessible to IN16B. However, extrapolations to lower or higher temperatures are also meaningful provided the system still obeys the Arrhenius law, and they will also be presented later in the manuscript.

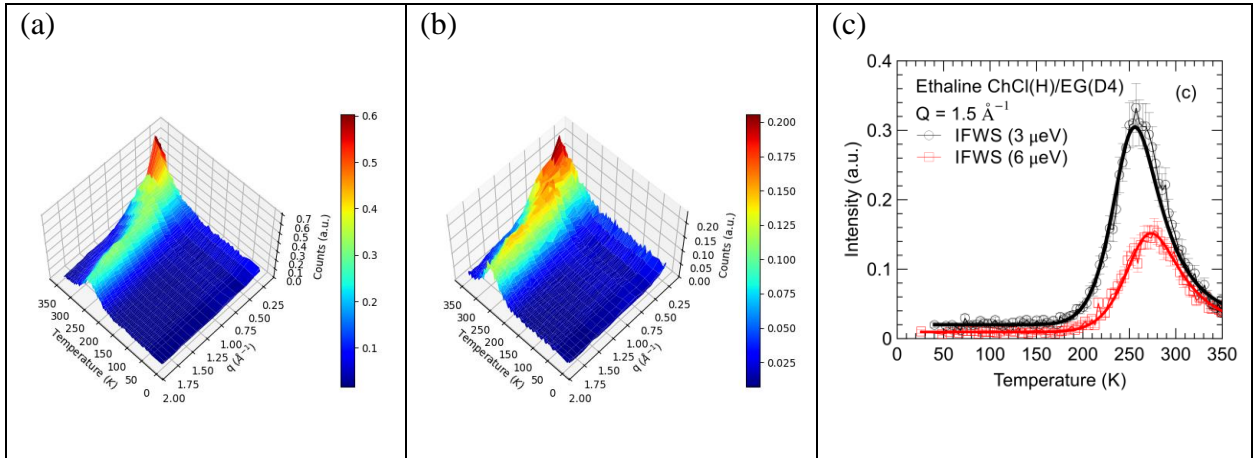

**FIG. 6.** Inelastic Fixed Window Scans of Ethaline ChCl(H)/EG(D4) obtained on IN16B for two different energy off-sets (a): 3 μeV (a), and (b): 6 μeV. The fitted model (solid line) is compared to the experimental intensity (symbols) at $Q = 1.5$ Å$^{-1}$ for the two energy off-sets (c)

## DISCUSSION

**Local molecular dynamics**

The local dynamics are illustrated by the dynamic structure factor $S_L(Q,\omega)$ comprising two terms that provide different information. The amplitude of the elastic term $A_L(Q)$ is determined by the entire spatial trajectory of the local motions, and as such, it contains geometrical but no temporal information on the dynamical properties. The latter information is contained in the width $\Gamma_L(Q)$ of the quasielastic broad component $L_L(Q,\omega)$, which is inversely proportional to the correlation time of the motion. The EISFs $A_L(Q)$ obtained from the fitting of Eq. (2) are illustrated in Fig. 7. The values decay from unity at low $Q$ to values that remain quite large (i.e., above 0.5) at $Q = 2$ Å$^{-1}$, which confirms that the trajectory associated to this fast motion is spatially restricted.

From a systematic trial of different classical models (e.g., Gaussian, rotation…), we came to the conclusion that the combination of a minimum of two different trajectories was required to reproduce the experimental EISFs. This is conceivable due to the level of complexity of the molecular systems under study, which comprise many inequivalent proton sites and various types of motions. The best fits were obtained for a model comprising two motions as described in the following.

For all systems, a Gaussian description of the localized motion of the entire molecule was adopted and it was applied to all the hydrogen atoms. This form is consistent with the localized motion of the molecule center of mass in a harmonic potential. It can also describe localized diffusive translational motion of molecule confined in a soft spherical environment.[37]

In addition, motions were considered for hydrogens present in specific sites, such as 3-sites jumps for the methyl groups of ChCl(H), and 2-sites jumps for the hydroxyl groups of ChCl and EG. The overall model writes as

$$A_L(Q) = \exp\left(\frac{-\langle r^2 \rangle Q^2}{6}\right)[(1 - x_{OH} - x_{Me}) + x_{OH}A_{OH}(Q) + x_{Me}A_{Me}(Q)] \quad (5)$$

where $\langle r^2 \rangle$ is the entire molecule mean squared displacement (msd) - i.e., twice the value of the mean squared offset of atomic positions from their equilibrium positions [26], $x_{Me}$ and $x_{OH}$ the molar fraction of methyl and hydroxyl hydrogens in the system, $A_{OH}(Q)$ and $A_{Me}(Q)$ the corresponding EISF represented by 2-sites and 3-site jumps, respectively. Note that the general form of Eq. 5 can be simplified to Eq. 6 for two systems (i.e., EG and Ethaline ChCl(D9)/EG(H)) that do not comprise methyl's hydrogens.

$$A_L(Q) = \exp\left(\frac{-\langle r^2 \rangle Q^2}{6}\right)[(1 - x_{OH}) + x_{OH}A_{OH}(Q)] \quad (6)$$

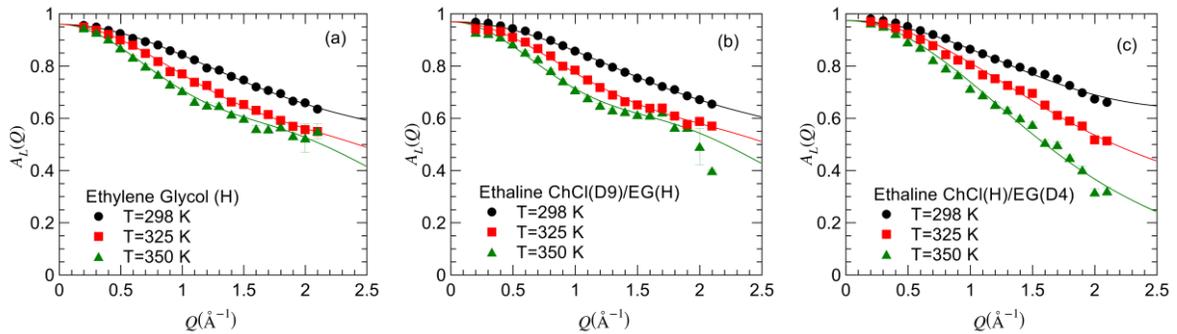

**FIG. 7.** Elastic incoherent structure factor $A_L(Q)$ of the local molecular motions (symbols) deduced from IN5B measurements at three different temperatures of (a) ethylene glycol, (b) ChCl(D9)/EG(H) and (c) ChCl(H)/EG(D4).

The parameters of the model are provided in Table S2. The radius of rotation the methyl hydrogens was fixed to the expected theoretical value. For EG and ChCl(D9)/EG(H), the jump length of hydroxyl H presented a similar systematic temperature dependence, which suggests that the amplitude of OH librational motion increases on increasing temperature. However, a different tendency was observed for ChCl(H)/EG(D4), as a possible consequence of the overwhelming contribution from methyl groups in this specific case.

In a previous QENS study of EG, Novikov et al. modelled the quasielastic scattering intensity of liquid EG by the combination of three motions: a Fickian translational diffusion and two molecular rotations occurring on different timescales.[35] An alternative description was also proposed by Sobolev et al.,[38] who fitted the entire dynamic structure factor by the Fourier transform of the stretched-exponential function. This phenomenological approach provides fit in good agreement with the experimental data, but lacks spatial information about the microscopic motions because no EISF is included in the model.

The best fits obtained with Eq. (5) are indicated as solid lines in Fig. 7. From a visual inspection of the EISFs, three important conclusions can already be drawn. First, $A_L(Q)$ systematically increases on decreasing the temperature. This indicates that the trajectory becomes more spatially restricted at low temperature, which can be easily understood. Second, the EISF of ethaline ChCl(D9)/EG(H) agrees very well with that of pure EG. In fact, as illustrated in Fig. 1(b), the scattering intensity of ethaline ChCl(D9)/EG(H) is largely dominated by the incoherent cross section of EG. The agreement between the EISFs of both systems demonstrates that the trajectory associated to the localized motions of EG in ethaline is actually the same as EG in the pure liquid state. This likely implies that the local environment around EG molecules, and specifically the H-bonded organization, is mainly preserved in the DES mixture of ethaline. In glyceline, it was similarly reported that the structural network remains dominated by H-bonded glycerol.[24] Third, it is striking from Fig. 7(c) that the EISF of ChCl(H)/EG(D4) decays faster with increasing $Q$ than for the two other systems.

We evaluated the amplitude of the confined molecular motion that is represented by the Gaussian term of Eq. 5 by $R = \sqrt{\langle r^2 \rangle}$, $\langle r^2 \rangle$ being the mean-squared-displacement determined from the EISF, as illustrated in Fig. S5. In fact, despite a different temperature dependance for ChCl(H)/EG(D4), the values of $R$ remained in a comparable range for the three systems and never exceeded 1 Å. In fact, we attribute the difference in EISF observed for ChCl(H)/EG(D4) to the additional intramolecular rotational motion performed by the nine hydrogen atoms of the three methyl groups.

In a previous QENS study of glyceline DES,[23,24] the EISF was related to the localized diffusive motions being confined within a transient cage of neighboring molecules, which precedes the cage-breaking long-range diffusion jumps. In line with our work, the authors observed in glyceline that the displacements associated with the localized motions were larger for ChCl than for glycerol. They additionally concluded on the faster local diffusive dynamics of ChCl. In this work, the authors emphasized that this observation is counterintuitive because ChCl is larger than glycerol, and as such, it actually presents slower self-diffusion coefficient than glycerol at the macroscopic scale. However, in the present study we did not observe difference in the width of the broad quasielastic line $\Gamma_L(Q)$ that would support this conclusion for ethaline. We actually found that $\Gamma_L(Q)$ ($\approx 0.2$ meV) was barely dependent on the transfer of momentum. The corresponding correlation time $\tau_L \approx 2 - 3$ ps presented a modest variation on the studied temperature range that is consistent with Arrhenius law. Somewhat slower dynamics were observed for ChCl(H)/EG(D4) (by a few percent), while indistinguishable relaxations times were obtained for both ChCl(D9)/EG(H) and pure EG. Accordingly, our interpretation of this dynamical process points to intramolecular molecular motions that exhibit larger amplitudes for ChCl than for EG. This can be simply rationalized due to the larger molecular size and, most importantly, to the motion of the peripheral location of the nine H atoms of the three methyl groups of ChCl. Additionally, our comparison between pure EG and ethaline ChCl(D9)/EG(H) demonstrates that the localized dynamics of EG are not significantly influenced by the presence of ChCl, despite the presence of hydrogen bonding interactions between the HBD and the ionic species.

**Translational diffusion. From the nanometer to macroscopic scale.**

The linewidth of the sharper Lorentzian quasielastic line $\Gamma_T$ found on IN5B at three distinct temperatures is illustrated in Fig. 8, as a function of the squared momentum transfer for the three studied systems (i.e., pure EG and ethaline samples with two different isotopic labelling). The values of $\Gamma_T$ were also evaluated for ethaline from the IFWS measured on IN16B Arrhenius law to reach the three temperatures measured on IN5B. These temperatures are far above the region (250 K - 285 K) around the maximum peak of IFWS. Indeed, the obtained values of $\Gamma_T$ up to 80 µeV, are one order of magnitude larger than the energy-offset used on IN16B. Despite this substantial extrapolation, excellent agreement between the two instruments operating on very different timescales is obtained.

At small to intermediate $Q$s, $\Gamma_T$ increases linearly with $Q^2$, which conforms the normal Fick's law of translational diffusion. At large $Q$s, $\Gamma_T$ bends and tends asymptotically towards a constant value denoted $1/\tau_0$ (Fig. 8). In fact, the Fickian diffusion model assumes a continuous motion process. Deviation from this assumption is observed when considering small displacements (*i.e.* for $Q$ larger than the inverse particle distance), where a discontinuous mechanism is related to the finite molecular size and the local order in the liquid. The linewidth was modelled by the well-known jump-diffusion model, which assumes that the translation motion proceeds by successive elementary jumps (solid lines in Fig. 8) [25]. Between two jumps, the particle remains localized for a typical residence time $\tau_0$ on a molecular site, with a spatial extension limited to the amplitude of local motions discussed in the previous sections. Applying the usual assumptions that the jump can be regarded as instantaneous with respect to the residence time $\tau_0$ spent by the particle on a site, and that the jump length $l$ is much larger than the spatial extension of each site, the linewidth of the Lorentzian was fitted with

$$\Gamma_T(Q) = \frac{D_T Q^2}{1 + \tau_0 D_T Q^2} \tag{6}$$

where $D_T$ is the diffusion coefficient and $\tau_0$ the mean residence time. Note that the deviation from Fickian dynamics decreased on heating, so that only the diffusion coefficient $D_T$ could be evaluated for pure EG and ChCl(D9)/EG(H) at 350K.

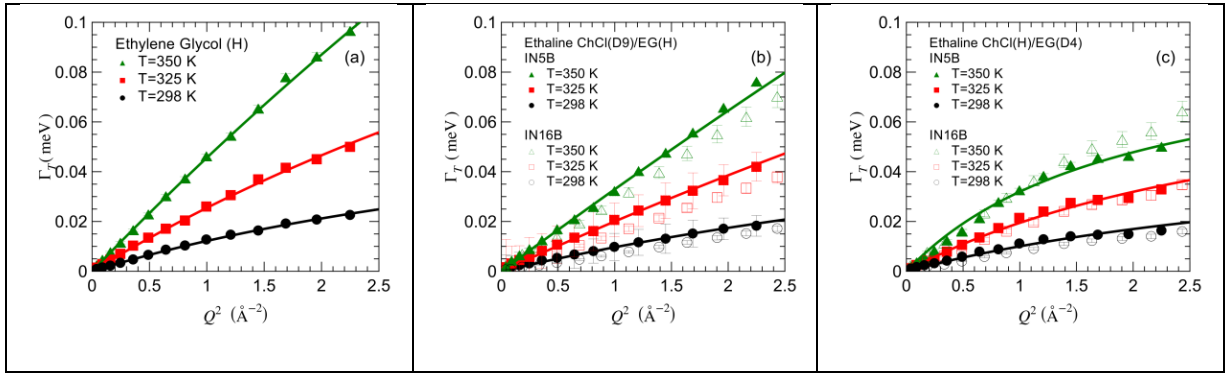

**FIG. 8**. Evolutions of the half width at half-maximum $\Gamma_T$ of the sharp Lorentzian attributed to translational diffusion as a function of $Q^2$ obtained from the fitting of QENS spectra measured on IN5B (filled symbols) and IN16B (open symbols) for Ethylene Glycol (left panel), Ethaline ChCl(D9)/EG(H) (middle panel), and Ethaline ChCl(H)/EG(D4) (right panel). Fit using the jump diffusion model (thin solid lines) at three temperatures, from top to bottom 350 K, 325 K, and 298 K (solid lines).

As seen in Fig. 8, the nanometer-scale translational dynamics of ethaline markedly depend on the isotopic composition of the DES. They are rather comparable to pure EG for ChCl(D9)/EG(H), but markedly slower for ChCl(H)/EG(D4). These differences are quantified by the values of the translational diffusion coefficient ($D_T$) and the residence time ($\tau_0$), which are presented in Arrhenius coordinates in Fig. 9 for the different systems.

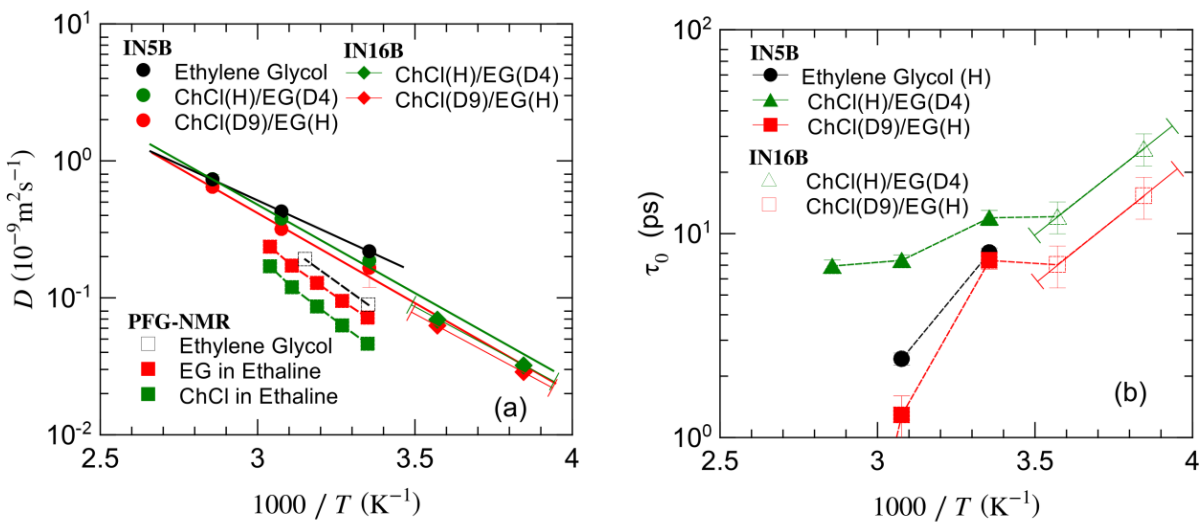

**FIG. 9**. (a) Translational diffusion coefficient $D_T$, (b) residence time $\tau_0$ of ethaline and ethylene glycol evaluated from the fit of IN5B spectra and IN16B IFWS. IN16B results are illustrated in the temperature range 250 – 285 K (|—|) around the maxima of the IFWS (symbols). In panel (a): the diffusion coefficients deduced from PFG NMR (square), and Arrhenius temperature fits of the translational diffusion coefficient (line).

The diffusion coefficients obtained from QENS are systematically larger than those derived from PFG-NMR measurements as shown in Fig. 9(a). Note that for pure EG, the diffusivities we obtained from QENS and PFG-NMR measurements are in quantitative agreement with the existing literature, as illustrated in Fig. S6. Moreover, the values of $D_T$ derived by QENS are nearly identical for the three systems, which is quite different from the observations made by PFG-NMR. These differences arise from the distinct time and length scales probed by each technique. Indeed, QENS examines translational short-range molecular diffusion over displacements of approximately 1 nm, whereas PFG NMR probes much larger displacements (~1 μm) that exceed considerably the size of molecular units. For EG, the observed discrepancy between the diffusivities measured on the macroscopic scale compared to the molecular motions at the molecular scale can be solely related to the extending H-bond network that reduces long-range transport. A similar effect can be also invoked for ethaline to explain why the diffusion coefficients obtained by NMR are smaller than those obtained by QENS. In addition, unlike pure liquid like EG, ethaline also shows another interesting feature: the relative reduction in the diffusivities measured by the two methods is different for the two ethaline constituents (ChCl and EG). On the microscopic scale, ChCl and EG have the same diffusivity, while at the macroscopic scale the diffusion ChCl is slower than that of EG. According to QENS, we observe that ChCl and EG molecules remain dynamically correlated on short time and length scales. This can be viewed as a direct effect of molecular associations between the DES components by strong hydrogen bonding interactions. This is in contrast with longer timescales as-probed by NMR, where molecular interactions are averaged out. In the latter case, the differences in translational diffusion between the DES constituents reflect differences in molecular size and solvent viscosity, as expected from classical hydrodynamics.

Unlike $D_T$, the residence time $\tau_0$ presented in Fig. 9(b) reveals a dependence on the isotopic composition of ethaline. The value of $\tau_0$ for ChCl(D9)/EG(H) is similar to pure EG, while it is systematically larger for ChCl(H)/EG(D4). We recall here that the cholinium ion is the major source of incoherent scattering for ChCl(H)/EG(D4), while it is EG for ChCl(D9)/EG(H). However, non-negligible incoherent scattering also arises from the partially deuterated component (33 % for ChCl(H)/EG(D4) and 17 % for ChCl(D9)/EG(H)) as illustrated in Fig. 1. Considering the dynamic structure factor as a weighed sum of two partial dynamic structure factors arising from the two compounds, the as-measured residence time $\tau_0$ can be effectively viewed as a weighted average quantity, that mostly arises from EG for ChCl(D9)/EG(H) and from ChCl for ChCl(H)/EG(D4). Therefore, we can conclude that the residence times of the two constituents of ethaline mixture are different, being longer for ChCl than for EG. On the quantitative level, due to contributions from the partially deuterated constituent, our work also indicates that the difference between the two constituents' dynamics is possibly even larger than the difference in residence time measured by QENS for the two different isotopic mixtures. A mean jump length $l$ was estimated by combining the diffusion coefficient and the residence time according to $l = \sqrt{\langle 6D_T\tau_0 \rangle}$. The obtained values of $l$ were in the typical range from 0.5 to 1 Å both for ChCl(D9)/EG(H) and for pure EG, while they were slightly larger for ChCl(H)/EG(D4) in the range from 0.7 to 1.7 Å.

The larger amplitude of ChCl jumps between adjacent basins that precede Fickian translational diffusion echoes the observation made in the previous part of the manuscript based on the EISF of the intra-basin local molecular motion. The two phenomena likely share a common origin that is related to the larger molecular size of ChCl compared to EG.

It is noteworthy that our observations reveal the existence of similarities but also significant differences in the dynamics of different prototypical choline chloride-based DES. At the macroscopic scale we observe that the diffusion coefficients of ChCl and the HBD components differ in a magnitude consistent with their respective hydrodynamic radii. This observation is in agreement with glyceline. [23,24]

A different situation is encountered at the nanometer-scale. In their QENS study on glyceline Wagle et al. [23] made distinction between the "long-ranged" Fickian diffusion that follows cage-breaking, and the faster localized diffusive motions confined in the transient cage of neighbor particles. In line with our work on ethaline, comparable diffusion coefficients were observed for glycerol and ChCl in glyceline, despite different hydrodynamic radii. This highlights how the H-bonded association between the HBD and ChCl induces a coupling of their nanoscale translation diffusion. This phenomenon should be relevant for other DES systems, given the predominance of strong intermolecular interactions and long-lived supramolecular units.

At shorter distances (large $Q$), we observed that the initial stages of translational diffusion in ethaline show transition from discontinuous jumps to continuous Fickian diffusion. Due to a limited $Q$-range, deviation from the Fickian behavior was not considered for glyceline. Therefore, the difference in residence time and mean jump length of EG and ChCl that we revealed for ethaline is unique, and we cannot conclude whether this observation can be transposed to other DES. At even shorter distances, however, differences appear when comparing glyceline and ethaline. Differential localized mobility of glyceline constituents has been linked to ChCl undergoing looser transient spatial confinement in the first coordination shell. While we observed that the amplitude of localized motion and the mean jump diffusion lengths were larger for ChCl than for EG, as a result of different molecular sizes, neither the corresponding correlation time $\tau_L \approx 2-3$ ps nor the residence time $\tau_0 \approx 8-30$ ps indicated that ChCl was actually the fastest component of ethaline. In contrast, we observed that the local dynamics of ChCl exhibit the same correlation time $\tau_L$ and a longer residence time $\tau_0$ than those of EG. This discrepancy between glyceline and ethaline dynamics could reflect the difference in molecular radius between glycerol and EG, the latter inducing a greater size asymmetry with respect to ChCl. An additional effect arises from the H-bond strengths and types of associations involving the two different HBD. It has been shown that the association between glycerol molecules in glyceline is stronger than or equal to that between glycerol and Cl [21]. According to Faraone et al [24], the resulting stability of the extended H-bond network of glycerol can explain its dominant role in determining the collective dynamics of the DES, while the local dynamics is rather determined by the more loosely bounded ChCl. The opposite situation was reported for ethaline by Zhang et al. [39] who found that the EG-Cl interactions were much stronger than those between EG molecules in ethaline.

Finally, it is noteworthy that the interpretation of our QENS results is consistent with recent deuteron NMR study by Hinz et al. [17,18]. Therein, the authors reported that the reorientation dynamics of ChCl was faster than the HBD in glyceline, while the opposite situation occurred for ethaline. This phenomenon was also rationalized by invoking the different H-bond propensities of the two HBD.

## CONCLUDING REMARKS

Nowadays, DES attract large interest as efficient solvents with tunable properties that are relevant for broad range of applications. Besides, they exhibit a number of unusual physical properties such as non-ideal mixing, supramolecular order and dynamic heterogeneity. These features do not only classify DES as fundamentally different from classical simple liquids, they confer to each DES characteristics that are composition-specific.

The question arises whether specific features observed for a given DES can be considered salient features of a whole class of similar systems and even considered universal properties of DES. It is therefore fundamentally needed to better understand the underlying origins of their uncommon physical properties.

In the present work, we address the question of the molecular dynamics of the two constituents of the prototypical DES ethaline. Combining PFG-NMR, time-of-flight, and backscattering QENS with isotopically labelled samples, we compared the different motions of ChCl and EG on a spatio-temporal range that extends from sub-nanometer to micrometer distances and from ps to ms times.

We confirm that the translational motions at the micrometer macroscopic conforms classical hydrodynamics. The composition dependences of the diffusivities of the two DES constituents reflect the differences in molecular size and viscosity, in line with expectations from Stoke-Einstein law. This is no longer the case at the nanoscale, where the comparable values of the diffusivities of ChCl and EG indicate long-term correlations between the two compounds. The shorter-time dynamics include molecular jump processes that precede Fickian diffusion, and localized motions that precede the breaking of the transient cage formed by neighboring molecules. The spatial extension of both motion trajectories is larger for ChCl than for EG, as a consequence of the larger molecular size of ChCl compared to EG. However, in contrast to glyceline, the local dynamics of ChCl is slower or equal to that of EG.

The difference between the short- and long-range diffusivities of ethaline, as well as the discrepancy between the dynamics of two apparently similar choline-based DESs, is striking. It reflects the predominance of H-bond interactions in determining dynamics at the nanoscale, in a manner that is difficult to predict from macroscopic properties. It is noteworthy that solid-liquid interfaces as well as nanoporous media are ubiquitous in technological applications of DES. As such, our findings underly the importance of deepening in a systematic manner our understanding of nanometer scale dynamics of DES in the bulk as well as in spatially confined geometries.

## SUPPLEMENTARY MATERIAL

See the supplementary material for neutron scattering cross sections of the samples, PFG-NMR diffusion measurements method, parameters extracted from the fits of IN5B spectra, and Arrhenius representation of the self-diffusion coefficient of pure EG.

## ACKNOWLEDGEMENTS

The experiments were conducted as part of the Ph.D. project of M. N. Kamar, who gratefully acknowledges funding from the University of Rennes. We express our gratitude to the Institut Laue-Langevin (ILL) for providing neutron beam time, which was central in the success of this study.

## AUTHOR DECLARATIONS

**Conflict of Interest**

The authors have no conflicts to disclose.

**Author Contributions**

**Mohammad Nadim Kamar:** Data Curation (equal); Formal analysis (lead); Investigation (equal); Methodology (equal); Visualization (lead); Writing – original draft (lead); Writing – review & editing (equal). **Armin Mozhdehei:** Investigation (equal); Writing – review & editing (equal). **Basma Dupont:** Investigation (equal); Writing – review & editing (equal). **Ronan Lefort:** Data Curation (equal); Investigation (equal); Project administration (equal); Supervision (equal); Writing – review & editing (equal). **Alain Moréac:** Investigation (equal); Supervision (secondary); Writing – review & editing (equal). **Jacques Ollivier:** Data Curation (equal); Investigation (equal); Software (equal); Writing – review & editing (equal). **Markus Appel:** Data Curation (equal); Investigation (equal); Software (equal); Writing – review & editing (equal). **Denis Morineau:** Conceptualization (lead); Funding acquisition (lead); Investigation (equal); Methodology (equal); Project administration (lead); Resources (equal); Supervision (lead); Visualization (equal); Writing – original draft (equal); Writing – review & editing (equal).

## DATA AVAILABILITY

Raw data generated at the Institut Laue-Langevin (ILL) large scale facility and the dataset will be publicly available after the embargo period (https://dx.doi.org/10.5291/ILL-DATA.6-07-112).[27] Derived data supporting the findings of this study are available from the corresponding author upon reasonable request.

# REFERENCES

[1] Y.-F. Lin, and I.-W. Sun, "Electrodeposition of zinc from a Lewis acidic zinc chloride-1-ethyl-3-methylimidazolium chloride molten salt," Electrochimica Acta **44**(16), 2771–2777 (1999).

[2] A.P. Abbott, K.E. Ttaib, G. Frisch, K.S. Ryder, and D. Weston, "The electrodeposition of silver composites using deep eutectic solvents," Phys. Chem. Chem. Phys. **14**(7), 2443 (2012).

[3] A. Sun, H. Zhao, and J. Zheng, "A novel hydrogen peroxide biosensor based on the Sn–ZnNPs/MWNTs nanocomposite film," Talanta **88**, 259–264 (2012).

[4] E. Gómez, P. Cojocaru, L. Magagnin, and E. Valles, "Electrodeposition of Co, Sm and SmCo from a Deep Eutectic Solvent," Journal of Electroanalytical Chemistry **658**(1–2), 18–24 (2011).

[5] M. Chirea, A. Freitas, B.S. Vasile, C. Ghitulica, C.M. Pereira, and F. Silva, "Gold Nanowire Networks: Synthesis, Characterization, and Catalytic Activity," Langmuir **27**(7), 3906–3913 (2011).

[6] Y. Xie, H. Dong, S. Zhang, X. Lu, and X. Ji, "Solubilities of $CO_2$, $CH_4$, $H_2$, CO and $N_2$ in choline chloride/urea," Green Energy & Environment **1**(3), 195–200 (2016).

[7] B.B. Hansen, S. Spittle, B. Chen, D. Poe, Y. Zhang, J.M. Klein, A. Horton, L. Adhikari, T. Zelovich, B.W. Doherty, B. Gurkan, E.J. Maginn, A. Ragauskas, M. Dadmun, T.A. Zawodzinski, G.A. Baker, M.E. Tuckerman, R.F. Savinell, and J.R. Sangoro, "Deep Eutectic Solvents: A Review of Fundamentals and Applications," Chem. Rev. **121**(3), 1232–1285 (2021).

[8] A. Jani, T. Sohier, and D. Morineau, "Phase behavior of aqueous solutions of ethaline deep eutectic solvent," Journal of Molecular Liquids **304**, 112701 (2020).

[9] N. Schaeffer, L.P. Silva, and J.A.P. Coutinho, "Comment on 'Structural Study of a Eutectic Solvent Reveals Hydrophobic Segregation and Lack of Hydrogen Bonding between the Components,'" ACS Sustainable Chem. Eng. **10**(27), 8669–8670 (2022).

[10] O.S. Hammond, D.T. Bowron, and K.J. Edler, "Liquid structure of the choline chloride-urea deep eutectic solvent (reline) from neutron diffraction and atomistic modelling," Green Chem. **18**(9), 2736–2744 (2016).

[11] L. Percevault, A. Jani, T. Sohier, L. Noirez, L. Paquin, F. Gauffre, and D. Morineau, "Do Deep Eutectic Solvents Form Uniform Mixtures Beyond Molecular Microheterogeneities?," J. Phys. Chem. B **124**(41), 9126–9135 (2020).

[12] S. Kaur, A. Gupta, and H.K. Kashyap, "Nanoscale Spatial Heterogeneity in Deep Eutectic Solvents," J. Phys. Chem. B **120**(27), 6712–6720 (2016).

[13] S. Kaur, and H.K. Kashyap, "Unusual Temperature Dependence of Nanoscale Structural Organization in Deep Eutectic Solvents," J. Phys. Chem. B **122**(20), 5242–5250 (2018).

[14] V. Alizadeh, D. Geller, F. Malberg, P.B. Sánchez, A. Padua, and B. Kirchner, "Strong Microheterogeneity in Novel Deep Eutectic Solvents," ChemPhysChem **20**(14), 1786–1792 (2019).

[15] C. D'Agostino, R.C. Harris, A.P. Abbott, L.F. Gladden, and M.D. Mantle, "Molecular motion and ion diffusion in choline chloride based deep eutectic solvents studied by 1H pulsed field gradient NMR spectroscopy," Phys. Chem. Chem. Phys. **13**(48), 21383 (2011).

[16] I. Delso, C. Lafuente, J. Muñoz-Embid, and M. Artal, "NMR study of choline chloride-based deep eutectic solvents," Journal of Molecular Liquids **290**, 111236 (2019).

[17] Y. Hinz, and R. Böhmer, "2H and 13C nuclear spin relaxation unravels dynamic heterogeneities in deep eutectic solvents of ethylene glycol, glycerol, or urea with choline chloride," The Journal of Chemical Physics **159**(22), 224502 (2023).

[18] Y. Hinz, and R. Böhmer, "Deuteron magnetic resonance study of glyceline deep eutectic solvents: Selective detection of choline and glycerol dynamics," The Journal of Chemical Physics **156**(19), 194506 (2022).

[19] D. Reuter, P. Münzner, C. Gainaru, P. Lunkenheimer, A. Loidl, and R. Böhmer, "Translational and reorientational dynamics in deep eutectic solvents," The Journal of Chemical Physics **154**(15), 154501 (2021).

[20] D. Reuter, C. Binder, P. Lunkenheimer, and A. Loidl, "Ionic conductivity of deep eutectic solvents: the role of orientational dynamics and glassy freezing," Phys. Chem. Chem. Phys. **21**(13), 6801–6809 (2019).

[21] S. Spittle, D. Poe, B. Doherty, C. Kolodziej, L. Heroux, M.A. Haque, H. Squire, T. Cosby, Y. Zhang, C. Fraenza, S. Bhattacharyya, M. Tyagi, J. Peng, R.A. Elgammal, T. Zawodzinski, M. Tuckerman, S. Greenbaum, B. Gurkan, C. Burda, M. Dadmun, E.J. Maginn, and J. Sangoro, "Evolution of microscopic heterogeneity and dynamics in choline chloride-based deep eutectic solvents," Nat Commun **13**(1), 219 (2022).

[22] H. Srinivasan, V.K. Sharma, V.G. Sakai, J.P. Embs, R. Mukhopadhyay, and S. Mitra, "Transport Mechanism of Acetamide in Deep Eutectic Solvents," J. Phys. Chem. B **124**(8), 1509–1520 (2020).

[23] D.V. Wagle, G.A. Baker, and E. Mamontov, "Differential Microscopic Mobility of Components within a Deep Eutectic Solvent," J. Phys. Chem. Lett. **6**(15), 2924–2928 (2015).

[24] A. Faraone, D.V. Wagle, G.A. Baker, E.C. Novak, M. Ohl, D. Reuter, P. Lunkenheimer, A. Loidl, and E. Mamontov, "Glycerol Hydrogen-Bonding Network Dominates Structure and Collective Dynamics in a Deep Eutectic Solvent," J. Phys. Chem. B **122**(3), 1261–1267 (2018).

[25] M. Bée, "Quasielastic Neutron Scattering," Principles and Applications in Solid State Chemistry, Biology and Materials Science, (1998).

[26] R. Mhanna, P. Catrou, S. Dutta, R. Lefort, I. Essafri, A. Ghoufi, M. Muthmann, M. Zamponi, B. Frick, and D. Morineau, "Dynamic Heterogeneities in Liquid Mixtures Confined in Nanopores," J. Phys. Chem. B **124**(15), 3152–3162 (2020).

[27] Morineau Denis, Appel Markus, Cristiglio Viviana, Kamar Mohammad Nadim, Lefort Ronan, Moréac Alain, Mozhdehei Armin, and Ollivier Jacques, "Deep Eutectic Solvents in nanopores: Insight into dynamic and structural heterogeneities. Institut Laue-Langevin (ILL) doi:10.5291/ILL-DATA.6-07-112," (2023).

[28] B. Frick, J. Combet, and L. Van Eijck, "New possibilities with inelastic fixed window scans and linear motor Doppler drives on high resolution neutron backscattering spectrometers," Nuclear Instruments and Methods in Physics Research Section A: Accelerators, Spectrometers, Detectors and Associated Equipment **669**, 7–13 (2012).

[29] M. Appel, and B. Frick, "Note: One order of magnitude better signal-to-noise ratio for neutron backscattering," Review of Scientific Instruments **88**(3), 036105 (2017).

[30] O. Arnold, J.C. Bilheux, J.M. Borreguero, A. Buts, S.I. Campbell, L. Chapon, M. Doucet, N. Draper, R. Ferraz Leal, M.A. Gigg, V.E. Lynch, A. Markvardsen, D.J. Mikkelson, R.L. Mikkelson, R. Miller, K. Palmen, P. Parker, G. Passos, T.G. Perring, P.F. Peterson, S. Ren, M.A. Reuter, A.T. Savici, J.W. Taylor, R.J. Taylor, R. Tolchenov, W. Zhou, and J. Zikovsky, "Mantid—Data analysis and visualization package for neutron scattering and μ SR experiments," Nuclear Instruments and Methods in Physics Research Section A: Accelerators, Spectrometers, Detectors and Associated Equipment **764**, 156–166 (2014).

[31] William M. Spees, Sheng-Kwei Song, Joel R, Garbow, Jeffrey J. Neil, and Joseph J. H. Ackerman, "Use of Ethylene Glycol to Evaluate Gradient Performance in Gradient-Intensive Diffusion MR Sequences," Magnetic Resonance in Medicine **68**, 319–324 (2012).

[32] I. Alfurayj, C.C. Fraenza, Y. Zhang, R. Pandian, S. Spittle, B. Hansen, W. Dean, B. Gurkan, R. Savinell, S. Greenbaum, E. Maginn, J. Sangoro, and C. Burda, "Solvation Dynamics of Wet Ethaline: Water is the Magic Component," J. Phys. Chem. B **125**(31), 8888–8901 (2021).

[33] W. Shi, X. Chen, and X. Wang, "Density and viscosity of choline chloride/ethylene glycol deep eutectic solvent based nanofluid," Journal of Molecular Liquids **395**, 123852 (2024).

[34] M. Zhong, Q.F. Tang, Y.W. Zhu, X.Y. Chen, and Z.J. Zhang, "An alternative electrolyte of deep eutectic solvent by choline chloride and ethylene glycol for wide temperature range supercapacitors," Journal of Power Sources **452**, 227847 (2020).

[35] A.G. Novikov, M.N. Robnikova, and O.V. Sobolev, "Reorientation and diffusion motions in liquid ethylene glycol," Physica B: Condensed Matter **350**(1–3), E363–E366 (2004).

[36] A. Jani, B. Malfait, and D. Morineau, "On the coupling between ionic conduction and dipolar relaxation in deep eutectic solvents: Influence of hydration and glassy dynamics," The Journal of Chemical Physics **154**, 164508 (2021).

[37] F. Volino, J.-C. Perrin, and S. Lyonnard, "Gaussian Model for Localized Translational Motion: Application to Incoherent Neutron Scattering," J. Phys. Chem. B **110**, 11217–11223 (2006).

[38] O. Sobolev, A. Novikov, J. Pieper, "Quasielastic neutron scattering and microscopic dynamics of liquid ethylene glycol" Chemical Physics **334** 36–44 (2007).

[39] Y. Zhang, D. Poe, L. Heroux, H. Squire, B.W. Doherty, Z. Long, M. Dadmun, B. Gurkan, M.E. Tuckerman, and E.J. Maginn, "Liquid Structure and Transport Properties of the Deep Eutectic Solvent Ethaline," J. Phys. Chem. B **124**(25), 5251–5264 (2020).